\documentstyle[aps,
preprint,
]{revtex}
\begin{document}

\title{Dark Soliton Excitations in Single Wall Carbon
       Nanotubes}
\author{Zheng-Mao Sheng$^{1,3}$
\thanks{Email address: zmsheng@mail.hz.zj.cn} and Guoxiang Huang$^{2,3}$}
\address{ $^1$Department of Physics, Zhejiang University, Hangzhou 310028, China\\
          $^2$Department of Physics and Key Laboratory for Optical and Magnetic
              Resonance Spectroscopy, East China Normal University, Shanghai 200062,
              China\\
          $^3$International Centre for Theoretical Physics, P.O. Box 586,
              I-34014 Trieste, Italy
        }

\date{\today}

\maketitle

\begin{abstract}

Dark soliton excitations are shown to exist in single wall carbon
nanotubes (SWCNTs). At first, the nonlinear effective interatomic
potential and the difference equation for longitudinal lattice
displacement are obtained for the SWCNTs by expanding Brenner's
many-body potential in a Taylor series up to fourth-order terms.
Then using a multi-scale method, for short wavelength lattice
excitations the equation of motion of lattice is reduced to the
cubic nonlinear Schr\"odinger equation. Finally, the dark soliton
solutions and relevant excitations in the SWCNTs with subsonic
velocity are discussed.

\end{abstract}
\pacs{ }


Carbon nanotube (CNT) is the name of ultrathin carbon fiber with
nanometer-size diameter and micrometer-size length and was
accidentally discovered by Sumio Iijima\cite{Iijima91} in 1991.
As a novel and potential carbon material, CNTs have received a
great deal of attention
\cite{Tanaka99,Tans98,Liu99,Kim99,Kong00}, and their many unique
properties, such as
structure\cite{Thess96,Journet97,Che98,Ren98,Fan99}, electronic
property \cite{Mintmire92,Hamada92,Saito92}, superconductivity
\cite{Tang01,Huang96,Benedict95,Sheng01} and elementary
excitations \cite{Sheng01,Lin97,Chamon00,Figge01,Astakhova01},
have been investigated. The structure of the CNT consists of an
enrolled graphite sheet, and can be classified into either
multi-wall or single wall CNT (MWCNT or SWCNT) depending on its
preparation method. The smallest SWCNT with a diameter of 4 $\AA$
was discovered in last year\cite{Iijima00,Wang00}.

In recent years, there has been considerable efforts for studying
the nonlinear localized excitations in
lattices\cite{sie,fla,hua1}. The nonlinear localized excitations
can transfer energy and be involved in various processes of
interest. There have been some speculations on the role played by
solitary excitations in heat transfer, polymer destruction, and
other processes occurring in molecular systems. The dynamical
properties of the CNTs are of great interest due to their
potential for useful practical applications \cite{Craighead00}.
The longitudinal soliton excitations governed by Korteweg-de Vries
(KdV) equation in the CNTs using Brenner's many-body potential
have been investigated in Ref.\cite{Astakhova01}. Such solitons
are obtained under the assumption of a long wavelength
approximation. It is known that, in one-dimensional (1D) lattices,
for a weekly nonlinear lattice excitation with a large spatial
extension, its amplitude (or envelope) is governed by the
nonlinear Schr\"{o}dinger (NLS) equation when the excitation is a
short wave wavepacket\cite{Tsurui72}. The nonlinear excitations in
this case are envelope solitons. It is interesting to consider the
envelope soliton excitations in the SWCNTs.

In this work we investigate the nonlinear effects in SWCNTs giving
rise to {\it dark} envelope solitons using Brenner's empirical
many-body potential for carbon systems\cite{Brenner90}. At first,
we introduce a quasi-one dimensional lattice model for armchair
(tube (m,m)\,) CNTs in an
 anharmonic approximation of
the simple analytical Brenner's potential. Then we derive a NLS
equation by using a multiple-scale approach. Finally, we discuss
the dark soliton solutions of the NLS equation and the physical
relevance of these excitations in the SWCNTs.

The SWCNTs are highly anisotropic and ordered objects. The surface of the
SWCNT is formed by a graphite sheet
folded into a cylinder with bond lengths and angles differing slightly
from graphite on account of
the strain induced by the folding. The interaction between adjacent
atoms $i$ and $j$ in the
SWCNTs can be described by the Brenner's many-body
potential\cite{Astakhova01}, which reads

\begin{equation}
E_{ij}^{b}=V_{ij}^{R} - {\bar B}_{ij} V_{ij}^{A},
 \label{pot}
\end{equation}
where $V_{ij}^{R}$ and $V_{ij}^{A}$ are, respectively, exponential
repulsive and attractive terms: $V_{ij}^{R}=27.27
\exp[-3.28(r_{ij}-1.39)]$ and $V_{ij}^{A}=33.27
\exp[-2.69(r_{ij}-1.39)]$. ${\bar B}_{ij} $ represents an
environment dependent many body coupling between atoms $i$ and
$j$ containing geometric information associated with the system.
The total energy of the system is obtained by summation of Eq.\
(\ref{pot}) over all bonds. The energy and distances are measured
in eV and angtr\"oms, respectively.

We take the cylindrical coordinate system $(R, \Phi,z)$ and align the
tube axis as the $z$ axis so that $2m$
atoms in the SWCNT layer have the same $z$ coordinates. The equilibrium
distance between layers, $l_0=1.26 \AA$
for a (5,5) tube.

Let us consider a cylindrically symmetrical disturbance in the
SWCNT geometry as done in  Ref.\ \cite{Astakhova01}, in which all
$2m$ atoms in the $n$th layer have identical $z$ and radical
displacements from their equilibrium positions $Z_{n}^{0}$ and
$R^{0}$: $Z_{n} = Z_{n}^{0} + \zeta_{n}$ and $R_{n} = R^{0} +
\rho_{n}$, where $Z_{n}$ and $R_{n}$ are the perturbed
coordinates and $ \zeta_{n}$ and $\rho_{n}$ are the displacements
from equilibrium positions. Then the coordinates of $i$th atom in
$n$th layer are $(R^{0} + \rho_{n}, \Phi_{i}^{0}, Z_{n}^{0} +
\zeta_{n})$ for weak nonlinear excitations the $\zeta$ and $\rho$
are much smaller than the characteristic length scale. So the
interatomic potential can be expanded in a Taylor series and the
terms up to fourth order retained:
\begin{equation}
E = E_0 + \sum_{n=1}^{N_l} 2m E_{n}
\label{En}
\end{equation}
where the $E_0$ is the ground state energy of relaxed SWCNT, $E_n
= \frac{1}{2}(E_1 + E_2 + E_3)$ is the energy of an atom in the
$n$th layer due to atomic displacements, where $ E_1$, $E_2$ and
$E_3$ are the energies of bonds emerging from any atom in the
$n$th layer.

We rewrite the Brenner's potential (\ref{pot}) by expanding it in
Taylor series up to the fourth-order terms as
\begin{eqnarray}
E(r) &=& \frac{A}{\sigma_2}\left(\sigma_2 \exp[-\sigma_1(r-r_0)]-
\sigma_1 \exp[-\sigma_2(r-r_0)]\right)
\nonumber \\
&=& \frac{f}{2} (r-r_0)^2 - \frac{f}{3 r_0} a_1 (r-r_0)^3 +
\frac{f}{4 r_0^2} a_2 (r-r_0)^4,
\label{Ur}
\end{eqnarray}
where $A=27.27$ eV, $\sigma_1=3.28 \AA^{-1}$, $\sigma_2 = 2.69
\AA^{-1}$ and $r_0=1.39 \AA$ and
$f=\sigma_1 A(\sigma_1 -\sigma_2)=52.77$eV$/\AA^2$,
$a_1=\frac{r_0}{2}(\sigma_1 + \sigma_2)=4.15$
and $a_2=\frac{r_0^2}{6}(\sigma_1^2 + \sigma_1\sigma_2 +
\sigma_2^2) = 8.64$.
Thus $E_i = E(r_i), i=1,2,3$ with
$$
r_1 = \{ ( R^0 + \rho_n)^2(\cos\Phi_{i+1}-\cos\Phi_{i})^2
+ (R^0 + \rho_n)^2 ( \sin\Phi_{i+1}-\sin\Phi_{i})^2 \}^{1/2}
= r_0 ( 1 + \rho_n/R^{0}),
$$
$$
r_2 = \{ r_0^2 + (\rho_{n+1} - \rho_n)^2 + (r_0^2-l_0^2)(\frac{\rho_{n+1}
+ \rho_n }{ R^0} +\frac{\rho_{n+1} \rho_n}{(R^0)^2}) + 2l_0 (\zeta_{n+1}
-\zeta_n)
+ (\zeta_{n+1}-\zeta_n)^2\}^{1/2} \\
$$
and $r_3= r_2(n+1 \to n-1)$.
Expanding $r_2$ and retaining the terms to the third-order, we get
$$
r_2= r_0 + X_1 + X_2 + X_3, $$
where
\begin{eqnarray}
X_1 &=& \frac{r_0^2 -l_0^2}{2 R^0 r_0}(\rho_{n+1} + \rho_n)
+ \frac{l_0}{r_0}(\zeta_{n+1}-\zeta_n),
\nonumber \\
X_2 &=& \frac{1}{2r_0}(\rho_{n+1} - \rho_n)^2
+ \frac{r_0^2-l_0^2}{2r_0(R^0)^2}\rho_{n+1}\rho_n
- \frac{(r_0^2-l_0^2)^2}{8r_0^3(R^0)^2} (\rho_{n+1} + \rho_n)^2
\nonumber\\
&&- \frac{(r_0^2-l_0^2)l_0}{2r_0^3R^0}(\rho_{n+1}
+ \rho_n)(\zeta_{n+1}-\zeta_n)
+ \frac{1}{2r_0}(1- \frac{l_0^2}{r_0^2})(\zeta_{n+1}-\zeta_n)^2, \\
\nonumber
X_3 &=& - \frac{1}{2r_0^3}\left( (\rho_{n+1} - \rho_n)^2
+ \frac{r_0^2 - l_0^2}{(R^0)^2}\rho_{n+1}\rho_n +(\zeta_{n+1}
-\zeta_n)^2\right ) \\
\nonumber
& &\cdot\left(\frac{ r_0^2 - l_0^2}{2R^0}(\rho_{n+1}+\rho_n)
+ l_0 (\zeta_{n+1}-\zeta_n)\right ).
\end{eqnarray}
Then we have
\begin{eqnarray}
E_n =&& \frac{r_0^2 f}{4 (R^0)^2}\rho_n^2
- \frac{r_0^2 f}{6 (R^0)^3}a_1 \rho_n^3
+ \frac{r_0^2 f}{8 (R^0)^4} a_2 \rho_n^4
\nonumber \\
&&+ \frac{1}{2} f X_1^2 + f X_1 X_2 - \frac{f a_1}{3 r_0}X_1^3
+ f ( \frac{1}{2}X_2^2 + X_1 X_3 -\frac{a_1}{r_0} X_1^2 X_2
+ \frac{a_2}{4r_0^2}X_1^4).
\label{En2}
\end{eqnarray}

If we neglect the radial degrees of freedom and reduce $\zeta$ to
the dimensionless unit $\zeta_n \to \zeta_n/l_0$, we get
\begin{equation}
E_n = \frac{b}{2}(\zeta_{n+1}-\zeta_n)^2
- \frac{b p}{3}(\zeta_{n+1}-\zeta_n)^3
+ \frac{b q}{4}(\zeta_{n+1}-\zeta_n)^4,
\label{En3}
\end{equation}
where $b = \frac{l_0^2}{r_0^2} l_0^2 f = 68.84 $eV,
$ p = -\frac{3}{2} + (\frac{3}{2} + a_1) \frac{l_0^2}{r_0^2} =3.14 $,
$~~ q = \frac{1}{2} -(2a_1 +3)\frac{l_0^2}{r_0^2} +(\frac{1}{2} + 2a_1
+ a_2)\frac{l_0^4}{r_0^4} = 2.99. $
The equation of motion for longitudinal displacements reads
\begin{eqnarray}
& & \frac{d^2 \zeta_n}{d t'^2}= [\zeta_{n+1} - 2\zeta_{n}
    + \zeta_{n-1}]
    - p [(\zeta_{n+1} - \zeta_{n})^2 -(\zeta_{n} - \zeta_{n-1})^2]
    \nonumber \\
& & \hspace{2cm} + q [(\zeta_{n+1} - \zeta_{n})^3 -(\zeta_{n}
    - \zeta_{n-1})^3],
    \label{Eq}
\end{eqnarray}
where we have made the dimensionless transformation for time $t$
as $t' =\sqrt{b/(M l_0^2)} t$,  M is the mass of carbon atom.

Now we investigate how the narrow band wave packets evolve by
nonlinear effect\cite{Tsurui72,Huang93}. We look for a solution
expanded in terms of a small but finite parameter $\varepsilon $
denoting the relative amplitude of the excitations:

\begin{equation}
\zeta_n = \sum_{\nu=0}^{\infty} \varepsilon^{\nu+1} u_n^{(\nu)}
\label{u1}
\end{equation}
and each $ u_n^{(\nu)}$ is also expanded in terms of harmonics
$exp(i l\theta_n)$ with "fast" variable $\theta_n = k n l_0 -
\omega t'$ as
\begin{equation}
u_n^{(\nu)} = \sum_{l= -\infty}^{\infty} u_{n,l}^{(\nu)}(\tau,
\xi_n)e^{i l\theta_n}. \label{unu}
\end{equation}
where  $\tau$ and $\xi_n$ are slowly vary variables defined by $
\tau = \varepsilon^2 t' $ and $\xi_n = \varepsilon (n l_0 -\lambda
t')$, respectively. $\lambda$ denotes the group velocity $\lambda
= \frac{\partial \omega}{\partial k}$ which will be determined
later. It is noted that the reality conditions
\begin{equation}
u_{n,l}^{(\nu)} = {u_{n,-l}^{(\nu)}}^{*} \label{re}
\end{equation}
should be satisfied for all $\nu$ and $l$.

Substituting Eqs. (\ref{u1}) and (\ref{unu}) into Eq.\ (\ref{Eq})
and equating the coefficients of various powers of $\varepsilon $
to zero, we get
\begin{equation}
\sum_{l} \{ -l^2 \omega^2 + 4 \sin^2\frac{l k l_0}{2}\}
u_{n,l}^{(0)} e^{il\theta_n} =0,
\end{equation}
\begin{eqnarray}
\sum_{l}&&\left\{ 2i(l\omega\lambda -l_0\sin{l k l_0})
\frac{\partial u_{n,l}^{(0)}}{\partial \xi_n} + (-l^2 \omega^2 + 4
sin^2\frac{l k l_0}{2}) \right\} u_{n,l}^{(1)} e^{il\theta_n}
\nonumber \\
& &= p \sum_{l}\sum_{l'} 8i\sin^2\frac{l k l_0}{2}\sin{l'kl_0}
u_{n,l}^{(0)} u_{n,l'}^{(0)} e^{i(l+l')\theta_n},
\end{eqnarray}
\begin{eqnarray}
\sum_{l}&&\{ (\lambda^2 - l_0^2 \cos{l k l_0})\frac{\partial^2
u_{n,l}^{(0)} }{\partial \xi_n^2} - 2il\omega\frac{\partial
u_{n,l}^{(0)} }{\partial \tau} +  2i(l\omega\lambda -l_0\sin{l k
l_0})\frac{\partial u_{n,l}^{(1)}}{\partial \xi_n}
\nonumber \\
 & &+ (-l^2 \omega^2
+ 4 sin^2\frac{l k l_0}{2} )u_{n,l}^{(2)}\} e^{il\theta_n}
 \nonumber \\
& &= p \sum_{l}\sum_{l'} \{ 8i\sin^2\frac{l k l_0}{2}\sin{l'kl_0}
( u_{n,l}^{(1)} u_{n,l'}^{(0)} + u_{n,l}^{(0)} u_{n,l'}^{(1)})
\nonumber \\
&  & + 8l_0\sin^2\frac{l k l_0}{2}\cos{l'kl_0} u_{n,l}^{(0)}
\frac{\partial u_{n,l'}^{(0)}}{\partial \xi_n} + 4 l_0 \sin{l k
l_0}\sin{l'kl_0}\frac{\partial u_{n,l}^{(0)}}{\partial\xi_n}
u_{n,l'}^{(0)} \}e^{i(l+l')\theta_n}
\nonumber  \\
&  & -q\sum_{l}\sum_{l'}\sum_{l''}4 \sin^2\frac{l k l_0}{2} \{
(e^{il'kl_0}-1)( e^{il''kl_0}-1) +  (e^{il'kl_0}-1)(1- e^{-il''kl_0})
\nonumber \\
&    &+(1- e^{-il'kl_0})(1- e^{-il''kl_0}) \}
u_{n,l}^{(0)}u_{n,l'}^{(0)} u_{n,l''}^{(0)}
e^{i(l+l'+l'')\theta_n}.
\end{eqnarray}
and the equations corresponding to higher order of $\varepsilon
$. To derive these equations, we have expanded the terms $
u_{n\pm 1,l}^{(\nu)}$ in Taylor series of $\varepsilon l_0$,
because $\xi_{n\pm 1}=\xi_n \pm \varepsilon l_0$.

From above equations, we obtain the dispersion relation $\omega^2
= 4\sin^2 \frac{k l_0}{2} $, here $k$ and $\omega$ being
respectively the wave number and the frequency of a lattice wave,
and the following equations for the envelopes of the lattice wave
\begin{equation}
u_{n,l}^{(0)} =0,  ~~~~~~ {\rm for} \,\, |l| \geq 2
\end{equation}
\begin{equation}
\frac{\partial^2 u_{n,0}^{(0)}}{\partial \xi_n^2}
 = \frac{8 p}{l_0}\frac{\partial}{\partial \xi_n}\vert u_{n,\pm 1}^{(0)}\vert^2, \label{u0}
\end{equation}
\begin{equation}
i \frac{\partial u_{n,1}^{(0)}}{\partial \tau}
 = \frac{l_0^2}{8}\omega \frac{\partial^2 u_{n,1}^{(0)}}{\partial \xi_n^2}
+ \omega \{ \frac{3}{2} q\omega^2 + p^2 (4-\omega^2)\} \vert
u_{n,1}^{(0)}\vert^2 u_{n,1}^{(0)} -p l_0\omega \frac{\partial
u_{n,0}^{(0)}}{\partial \xi_n} u_{n,1}^{(0)}.
\label{NLS}
\end{equation}
The Eq.\ (\ref{u0}) is easily integrated once and written as
\begin{equation}
\frac{\partial u_{n,0}^{(0)}}{\partial \xi_n}
 = \frac{8 p}{l_0}\left\{ \vert u_{n,1}^{(0)}\vert^2 -C \right\},
\label{u01}
\end{equation}
where the $-\frac{8 p}{l_0}C $ is an integration constant.
Therefore from Eq.\ (\ref{NLS}) we obtain the single equation to
determine $ u_{n,1}^{(0)}$
\begin{equation}
i \frac{\partial u_{n,1}^{(0)}}{\partial \tau}
 = \frac{l_0^2}{8}\omega \frac{\partial^2 u_{n,1}^{(0)}}{\partial \xi_n^2}
+ \omega \left\{ \frac{3}{2} q\omega^2 - p^2 (4+\omega^2)\right\}
\vert u_{n,1}^{(0)}\vert^2 u_{n,1}^{(0)} + 8 p^2\omega C
u_{n,1}^{(0)}.
\label{NLS1}
\end{equation}
By replacements of variables, we obtain the standard cubic nonlinear
Schr\"odinger equation
\begin{equation}
i \frac{\partial \phi}{\partial z}
 = \frac{1}{2} \frac{\partial^2 \phi}{\partial x^2}
- \kappa \vert \phi\vert^2 \phi
\label{NLS2}
\end{equation}
where $\phi = \exp(i8p^2\omega\int Cd\tau) u_{n,1}^{(0)}$,
$z=\omega \tau$, $x=\frac{2}{l_0}\xi_n$ and
\begin{equation}
\kappa=(p^2 - \frac{3}{2}q)\omega^2 + 4 p^2.
\label{ka}
\end{equation}

Here $\kappa$ is always positive,  because $p^2 - \frac{3}{2}q = 5.37 >0 $.
It is known that the reversal of
the sign of $\kappa$ leads not only to a change in the physics picture of
the phenomena described
by Eq.\ (\ref{NLS2}), but
also to a considerable restructuring of the mathematical formalism necessary
for its solution.

Now we investigate the modulational stability and
solitons\cite{Kivshar98,Huang01}. The NLS Eq.\ (\ref{NLS2}) has
the simplest solution in the form of a continuous wave given by
the expression,
\begin{equation}
\phi = \phi_0 e^{i \alpha x + i\beta z},   ~~~~~~ \beta =
(\frac{1}{2} \alpha^2 + \kappa \phi_0^2) \label{s0}
\end{equation}
where $\phi_0$ is a constant. Let us investigate the linear stability
of the exact solution, Eq.\ (\ref{s0}),
against small perturbation as
\begin{equation}
\phi = (\phi_0 +\chi) e^{i (\alpha x +\beta z + \psi)}, \label{s1}
\end{equation}
where the function $\chi$ and derivative of the phase $\psi$ are
assumed to be small. Substituting Eq.\ (\ref{s1})
into Eq.\ (\ref{NLS2}), we look for the solution to these functions
in the form, $\chi$, $\psi \sim
\exp(i\Omega z - i Q x)$, then obtain the dispersion relation as
\begin{equation}
Q^2(\kappa \phi_0^2 + \frac{1}{4} Q^2) = (\alpha Q + \Omega)^2
\end{equation}
which means that the small excitation are always stable when
$\kappa > 0$ and the small amplitude waves can propagate along
the background. As a result, Eq.\ (\ref{NLS2}) has soliton
solutions in the form of localized "dark" pulses created on the
continuous wave background. The NLS equation with the boundary
conditions $\vert \phi\vert \to \phi_0,$ at $x \to \pm\infty,$ is
exactly integrable by the inverse scattering
method\cite{Zakharov73},
\begin{equation}
\phi = \phi_0(\cos\varphi\tanh\Theta + i\sin\varphi)e^{i \alpha x
+ i\beta z}, \label{s2}
\end{equation}
with $\Theta = \sqrt{\kappa}\phi_0\cos{\varphi} [ x - x_0 + (
\alpha + \sqrt{\kappa}\phi_0\sin{\varphi})z]$, where $\phi_0 $, $
\varphi $, $ \alpha $ and $x_0$ are free parameters. The $\phi_0$
is the amplitude, and $\varphi $ corresponds to the total phase
shift across the dark soliton, $2\varphi$. Then we obtain
\begin{eqnarray}
u_{n,1}^{(0)} &=& \exp(-i8p^2\omega\int Cd\tau) \phi
\nonumber \\
u_{n,0}^{(0)}&=&\frac{4 p}{\sqrt{\kappa}\phi_0\cos\varphi } \left
[ (\phi_0^2 -C)\Theta -\phi_0^2 \cos^2\varphi \tanh\Theta \right ]
\end{eqnarray}
If we take $C=\phi_0^2$ for simplicity, then we obtain dark
soliton excitations for the longitudinal displacement $\zeta_n$
up to the lowest order as
\begin{equation}
\zeta_n = A_0 \tanh\Theta [\cos X -\frac{2p}{\sqrt{\kappa}} ]
-A_0 \tan\varphi \sin X
\end{equation}
with $$ X=(k+\frac{2}{l_0}\alpha')nl_0 -\left
[2\alpha'\cos\frac{kl_0}{2} + ( 2-\alpha'^2 - \frac{\kappa
A_0^2}{2\cos^2\varphi}+\frac{4
p^2A_0^2}{\cos^2\varphi})\sin\frac{kl_0}{2}
\right]\sqrt{\frac{b}{Ml_0^2}}t, $$ $$
\Theta=\frac{\sqrt{\kappa}}{l_0}A_0 [nl_0-(\cos\frac{k
l_0}{2}-(\alpha'+\frac{\sqrt{\kappa}}{2}A_0\tan\varphi)\sin\frac{k
l_0}{2})\sqrt{\frac{b}{M}}t], $$ where
$A_0=2\phi_0\varepsilon\cos\varphi$ is the rescaling amplitude
and $\alpha'=\varepsilon\alpha$ is the rescaling parameter.

The velocity of the dark soliton is given by $ v_g= [\cos\frac{k
l_0}{2}-(
\alpha'+\frac{\sqrt{\kappa}}{2}A_0\tan\varphi)\sin\frac{k
l_0}{2}]v_{sound}$, where $ v_{sound}=\sqrt{b/M} = 22 $km/sec is
the longitudinal sound velocity, which means the the velocity of
dark soliton in the SWCNT is subsonic.

To summarize, We have obtained the nonlinear effective interatomic
potential for the SWCNTs by expanding Brenner's many-body
potential in a Taylor series up to the fourth-order terms and the
difference equation for longitudinal lattice displacements. Using
the multi-scale method, for short wavelength excitations the
equation of motion of lattice has been reduced to the cubic
nonlinear Schr\"odinger equation. We have discussed also the
envelope dark soliton excitations in the SWCNTS. The dark soliton
is the excitation of a dip excited on the continuous background of
a high frequency harmonic oscillation of longitudinal
displacement, and its velocity is subsonic. We note that the KdV
soliton obtained in Ref.\ \cite{Astakhova01} is a weak nonlinear
excitation valid only in a long wavelength approximation with a
supersonic propagating velocity.

The authors would like to warmly thank the hospitality and support received
at the Abdus Salam International Centre for
Theoretical Physics, Trieste, Italy, where this work was completed.
Sheng thanks S.Y. Lou for useful discussion.
This work is also supported in part by the funds from Pandeng Project of
China, NSFC  and Zhejiang Provincial Natural
Science Foundation of China.


\end{document}